\documentclass[prd,twocolumn,superscriptaddress,amssymb,amsfonts,amsmath,nofootinbib,preprintnumbers]{revtex4-2}
\usepackage{graphicx,slashed,xcolor,multirow,bbold,mathtools,sidecap,tikz,bm,ulem,enumitem,booktabs,array,amsmath,amssymb}
\usepackage[colorlinks=true,linktoc=page,linkcolor=purple,citecolor=teal,urlcolor=magenta]{hyperref}
\usepackage[compat=1.1.0]{tikz-feynman}
\usepackage{siunitx,adjustbox,comment}
\definecolor{lime}{HTML}{A6CE39}
\DeclareRobustCommand{\orcidicon}{\hspace{-2.1mm}
\begin{tikzpicture}
\draw[lime,fill=lime] (0,0.0) circle [radius=0.13] node[white] {{\fontfamily{qag}\selectfont \tiny ID}}; \draw[white,fill=white] (-0.0525,0.095) circle [radius=0.007]; 
\end{tikzpicture} \hspace{-3.5mm} }
\foreach \x in {A, ..., Z} {\expandafter\xdef\csname orcid\x\endcsname{\noexpand\href{https://orcid.org/\csname orcidauthor\x\endcsname} {\noexpand\orcidicon}}}

\let\emph\textit

\begin{document}

\preprint{TTP25-028, P3H-25-059, ZU-TH 53/25, ICPP-97}

\title{Searching for a Charged Higgs Boson in Top-Quark Decays via the $WZ$ Mode}

\author{Saiyad Ashanujjaman\orcidA{}}
\email{saiyad.ashanujjaman@kit.edu}
\affiliation{Institut f\"ur Theoretische Teilchenphysik, Karlsruhe Institute of Technology, Engesserstra\ss e 7, D-76128 Karlsruhe, Germany}
\affiliation{Institut f\"ur Astroteilchenphysik, Karlsruhe Institute of Technology, Hermann-von-Helmholtz-Platz 1, D-76344 Eggenstein-Leopoldshafen, Germany}

\author{Andreas Crivellin\orcidB{}}
\email{andreas.crivellin@icrea.cat}
\affiliation{Universitat Autònoma de Barcelona, 08193 Bellaterra, Barcelona.}
\affiliation{ICREA, Instituci\'o Catalana de Recerca i Estudis Avan\c{c}ats,
Passeig de Llu\'{\i}s Companys 23, 08010 Barcelona, Spain.}

\author{Siddharth P.~Maharathy\orcidC{}}
\email{siddharth.prasad.maharathy@cern.ch}
\affiliation{School of Physics and Institute for Collider Particle Physics, University of the Witwatersrand, Johannesburg, Wits 2050, South Africa}
\affiliation{iThemba LABS, National Research Foundation, PO Box 722, Somerset West 7129, South Africa}
\affiliation{Indian Institute of Science Education and Research Pune, Dr.~Homi Bhabha Road, Pune 411008, India}

\author{Bruce Mellado\orcidD{}}
\email{bmellado@mail.cern.ch}
\affiliation{School of Physics and Institute for Collider Particle Physics, University of the Witwatersrand,
Johannesburg, Wits 2050, South Africa}
\affiliation{iThemba LABS, National Research Foundation, PO Box 722, Somerset West 7129, South Africa}

\begin{abstract}
Top-quark decays are sensitive probes of light charged Higgs bosons ($H^\pm$) due to the sizable $t\bar t$ production cross section at the LHC in conjunction with their distinct experimental signatures. While dedicated ATLAS and CMS searches considered only $H^\pm$ decays into $\tau\nu$, $cs$, or $cb$ for $m_{H^\pm}<m_t$, the $WZ$ channel remains unexplored, despite being the dominant mode in $SU(2)_L$ triplet models. Since, top-quark pair production with $t \to H^\pm b$ and $H^\pm \to WZ$ gives rise to $t\bar{t}Z$-like signatures, we recast existing $t\bar{t}Z$ analyses to search for signs of charged Higgs bosons and set novel limits on the product of branching fractions Br$(t\to H^\pm b) \times $Br$(H^\pm\to WZ)$. These constraints turn out to be at the sub-permille level, despite the observed $2\sigma$ preference for a non-zero value. Interpreted within the Georgi--Machacek model, these constraints probe the $Y=0$ triplet vacuum expectation value at the ${\cal O}(10)\,$GeV level, providing stronger sensitivity than the existing $cs$ and $\tau\nu$ searches while remaining compatible with electroweak precision data. Moreover, the $2\sigma$ preference for a non-zero cross section further strengthens the cumulative case for a $\approx152$\,GeV boson as suggested, in particular, by di-photon excesses.
\end{abstract}

\maketitle

\section{Introduction}
With the discovery of the Brout-Englert-Higgs boson~\cite{Higgs:1964ia, Englert:1964et, Higgs:1964pj, Guralnik:1964eu, Higgs:1966ev, Kibble:1967sv} at the LHC~\cite{Aad:2012tfa, Chatrchyan:2012ufa}, the particle content of the Standard Model (SM) has been confirmed experimentally. Although the measured properties of this $125$\,GeV Higgs are consistent with the SM expectations~\cite{Langford:2021osp, ATLAS:2021vrm}, this does not exclude the existence of additional scalar bosons provided their role in electroweak symmetry breaking is small. In fact, a plethora of extensions of the SM Higgs sector have been proposed, introducing $SU(2)_L$ singlets~\cite{Silveira:1985rk, Pietroni:1992in, McDonald:1993ex}, doublets~\cite{Lee:1973iz, Fayet:1974pd, Fayet:1977yc, Haber:1984rc, Kim:1986ax, Peccei:1977hh, Turok:1990zg}, and triplets~\cite{Ross:1975fq, Konetschny:1977bn, Cheng:1980qt, Lazarides:1980nt, Schechter:1980gr, Magg:1980ut, Mohapatra:1980yp, Gunion:1989ci, Rizzo:1990uu, Chardonnet:1993wd, Blank:1997qa}, and even higher representations.

The search for such new Higgs bosons is a major component of the LHC program, resulting in many dedicated analyses~\cite{ATLAS:2024itc, Liu:2024sgc}. In particular, charged Higgses are probed via several production mechanisms and decay channels~\cite{ATLAS:2024itc, Horii:2024qiw}. For masses smaller than the top-quark ($m_{H^\pm}<m_t$), they can be produced from top decays, $t \to H^\pm b$~\cite{Rizzo:1989ci, CoarasaPerez:1998sqz, Bejar:2000ub}---a promising avenue given the large $t\bar t$ production cross section at the LHC and the distinctive high-multiplicity final states involving leptons and ($b$-)jets. For this production mechanism, searches have used the $cs$~\cite{CMS:2020osd, ATLAS:2024oqu}, $cb$~\cite{ATLAS:2023bzb}, and $\tau\nu$~\cite{CMS:2019bfg, ATLAS:2024hya} decay modes of $H^\pm$. The corresponding upper limits on the product of branching fractions ${\rm Br}(t \to H^\pm b) \times {\rm Br}(H^\pm \to XY)$ ranges from $0.47\%$--$0.11\%$ ($XY=cs$), $0.15\%$--$0.42\%$ ($XY=cb$), and $0.16\%$--$0.02\%$ ($XY=\tau\nu$) at 95\% confidence level (CL) for $m_{H^\pm}$ between 100\,GeV and 160\,GeV.{\footnote{Notably, in the $cb$ channel, a moderate excess with a global significance of $2.5\sigma$ has been observed near a mass of $130\,$GeV, which, however points towards a non-minimal flavour structure~\cite{Crivellin:2023sig, Coloretti:2025yji}.} Moreover, LEP experiments set a lower limit of about 80\,GeV on $m_{H^\pm}$ for $cs$ and/or $\tau\nu$ decays~\cite{LEPHiggsWorkingGroupforHiggsbosonsearches:2001ogs}.

In contrast, the decay $H^\pm \to WZ$ (denoting both off-shell cases $H^\pm \to W^*Z$ and $H^\pm \to WZ^*$) has not been subjected to any dedicated ATLAS or CMS searches in the low mass region ($m_{H^\pm}<m_t$). While being loop-induced in $SU(2)_L$-doublet models, it can be the dominant decay mode in $SU(2)_L$-triplet scenarios~\cite{Georgi:1985nv, Chanowitz:1985ug, Gunion:1989ci, Cheung:1994rp, Cheung:2002gd, Asakawa:2005gv}. In this work, we focus on charged Higgs bosons with masses between 100\,GeV and 160\,GeV, produced in top-quark decays and subsequently decaying via $H^\pm \to WZ$ (see Fig.~\ref{fig:Feynman})~\cite{DiazCruz:1999mq}. This shares its experimental signatures with $t\bar t Z$ and $tWZ$ production, namely final states with $b$-jets and three or four leptons. Therefore, we can constrain charged Higgs bosons by recasting existing $t\bar tZ$ analyses done within the SM context~\cite{ATLAS:2023eld, CMS:2024mke}.

Note that this mass region is particularly interesting in light of the indications for a new Higgs boson at 152\,GeV in associated di-photon production~\cite{Crivellin:2021ubm,Bhattacharya:2023lmu, Bhattacharya:2025rfr} and multi-lepton final states~\cite{vonBuddenbrock:2016rmr,vonBuddenbrock:2017gvy,Buddenbrock:2019tua,vonBuddenbrock:2020ter,Coloretti:2023wng}. Furthermore, the $SU(2)_L$ triplet is a prime candidate to be involved in the explanation of these anomalies~\cite{Ashanujjaman:2024pky, Crivellin:2024uhc, Ashanujjaman:2024lnr, Banik:2023vxa,Coloretti:2023yyq}. In particular, in the Georgi--Machacek (GM) model~\cite{Georgi:1985nv}, the $Y=0$ and $Y=1$ triplet sectors have equal vacuum expectation values (VEVs), allowing sizable triplet VEVs while remaining compatible with electroweak precision data. Furthermore, in the generic GM model~\cite{Ashanujjaman:2025puu}, where the requirement of custodial symmetry in the scalar potential is relaxed, the relevant charged and neutral states can, in the limit of small mixing between the two triplet sectors, originate predominantly from the $Y=0$ triplet, with the charged state decaying dominantly into $WZ$.

\begin{figure}[htb!]
\centering
\begin{tikzpicture}[baseline=(current bounding box.center)]
\begin{feynman}
% interaction point
\vertex (a);
% gluons from protons
\vertex [above left=1.2cm of a] (c);
\vertex [above=0.1cm of c] (cu); 
\vertex [below=0.1cm of c] (cd);

\vertex [below left=1.2cm of a] (d); 
\vertex [above=0.1cm of d] (du); 
\vertex [below=0.1cm of d] (dd);
% incoming protons
\vertex [left=1.2cm of c] (p1) {$p$}; 
\vertex [left=1.2cm of cu] (p1u); 
\vertex [left=1.2cm of cd] (p1d);

\vertex [left=1.2cm of d] (p2) {$p$}; 
\vertex [left=1.2cm of du] (p2u); 
\vertex [left=1.2cm of dd] (p2d);

% outgoing protons
\vertex [above right=1.2cm of c] (pp1) {$p$}; 
\vertex [above right=1.15cm of cu] (pp1u); 
\vertex [above right=1.26cm of cd] (pp1d); 

\vertex [below right=1.2cm of d] (pp2) {$p$}; 
\vertex [below right=1.26cm of du] (pp2u); 
\vertex [below right=1.14cm of dd] (pp2d); 

% central propagator
\vertex [right=1.3cm of a] (b) ;

% outgoing top pair
\vertex [above right=1.3cm of b] (e);
\vertex [below right=1.3cm of b] (f);

% H+ branch
\vertex [above right=0.7cm of e] (j); 
\vertex [above right=0.7cm of j] (m){$W^\pm$};
\vertex [below right=0.7cm of j] (n){$Z$};
\vertex [below right=0.7cm of e] (i) {$\bar b$};

% other top decay
\vertex [above right=0.7cm of f] (k) {$W$};
\vertex [below right=0.7cm of f] (l) {$b$};

\diagram{
% protons to gluons
(p1) -- [fermion] (c) -- [gluon] (a), (p1u) -- (cu), (p1d) --  (cd),
(p2) -- [fermion] (d) -- [gluon] (a),  (p2u) -- (du), (p2d) -- (dd),

% gluons to protons
(c) -- [fermion] (pp1), (cu) -- (pp1u), (cd) -- (pp1d),
(d) -- [fermion] (pp2), (du) -- (pp2u), (dd) -- (pp2d),

% gluon fusion
(a) -- [gluon, edge label=$g$] (b),

% top pair
(f) -- [fermion, edge label=$t$] 
(b) -- [fermion, edge label=$\bar t$] (e),

% top -> H+ b
(j) -- [scalar, edge label'=$H^\pm$] (e) -- [fermion] (i),
(l) -- [fermion] (f) -- [boson] (k),

% H+ decay
(j) -- [boson] (m),
(j) -- [boson] (n)
};
\end{feynman}
\node[draw, circle, minimum size=12pt, inner sep=0pt, fill=gray!40] at (c) {};
\node[draw, circle, minimum size=12pt, inner sep=0pt, fill=gray!40] at (d) {};
\end{tikzpicture}
\caption{Representative Feynman diagram for $pp \to t\bar{t}$ with $t \to H^\pm b$ and $H^\pm \to W^\pm Z$, leading to a $t\bar{t}Z$-like signature.}
\label{fig:Feynman}
\end{figure}
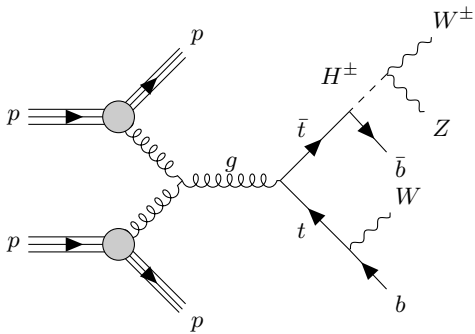

\section{Analyses of $t\bar{t}Z$ differential distributions}
We consider a charged Higgs boson produced via top-quark decay at the LHC. Thus, its production cross section from top-quark decays is approximately
\begin{align}
\sigma(H^\pm\!+2b+W)\approx 2\sigma(pp\to t\bar t) \times {\rm Br}(t\to H^\pm b) 
%\times \{1-{\rm Br}(t\to H^\pm b)\}
\,,
\end{align}
with $\sigma(pp\to t\bar t)= 832^{+46}_{-51}$\,fb for $m_t = 172.5$\,GeV at the 13\,TeV LHC within the SM~\cite{Czakon:2011xx}.\footnote{Note that the cross section for $t\bar t$ production with both top quarks decaying to $H^\pm b$ is negligible, as seen from the bounds we later obtain on Br$(t\to H^\pm b)$.} We then consider the decay $H^\pm \to WZ$, where one of the vector bosons is off-shell.\footnote{The main contribution to the signal region originates from $W^*Z$. However, we also included $WZ^*$ in our simulation since the ratio of the two modes can be calculated model-independently.} This $t\bar t Z$-like signature enables us to use the measurements of differential $t\bar tZ$ and $tWZ$ cross sections by CMS~\cite{CMS:2024mke} and ATLAS~\cite{ATLAS:2023eld}. 

The CMS analysis provides differential cross sections for the sum of $t\bar tZ$ and $tWZ$ production (within the SM), unfolded to the parton level (after radiation but before hadronization), as functions of the $Z$-boson transverse momentum ($p_T(Z)$), the transverse momentum of the lepton from the $W$ boson, ($p_T(\ell_W)$), the azimuthal angle between the two $Z$ leptons ($\Delta\phi(\ell^+,\ell^-)$), the angular separation between the $Z$ boson and the $W$-lepton ($\Delta R(Z,\ell_W)$), and the cosine of the angle between the $Z$ boson and the negatively charged lepton ($\cos\theta^*_Z$). The ATLAS analysis, on the other hand, reports $t\bar tZ$ differential cross sections unfolded to both particle and parton levels covering 15 observables (see Table~15 of Ref.~\cite{ATLAS:2023eld}).

For the validation of our setup, we simulate the SM processes $pp \to t\bar tZ$ and $tWZ$ using {\tt MadGraph5\_aMC\_v3.5.3}~\cite{Alwall:2014hca, Frederix:2018nkq} with the {\tt NNPDF31\_nlo\_as\_0118\_1000} parton distribution function~\cite{NNPDF:2017mvq} at next-to-leading order (NLO) accuracy in QCD\footnote{At NLO, $tWZ$ production interferes with the leading order $ttZ$ process. We use the {\tt MadSTR} plugin~\cite{Frixione:2019fxg}, which removes overlap at the amplitude level using the diagram removal approach.} and compared them to the SM predictions provided by the experimental collaborations, which are based on MadGraph5~\cite{Alwall:2014hca, Frederix:2018nkq}, {\tt PYTHIA}~\cite{Bierlich:2022pfr} and in the case of ATLAS also {\tt Herwig}~\cite{Bellm:2015jjp}. The obtained parton-level events are interfaced with {\tt Pythia 8.3}~\cite{Sjostrand:2014zea} containing the {\tt CMS-CUETP8S1-CTEQ6L1} tune~\cite{CMS:2015wcf} to model particle decays, parton showering, and radiation. The new physics (NP) signal process $pp \to t\bar t \to W^\mp bH^\pm b$ is simulated analogously for 22 benchmark values of $m_{H^\pm}$ in the 100\,GeV--160\,GeV range.
For the reconstruction and selection of physics objects, namely leptons (electrons and muons) and jets (including $b$-tagged jets), we closely follow the respective CMS and ATLAS analyses. Jets are clustered with the anti-$k_T$ algorithm~\cite{Cacciari:2008gp} implemented in {\tt FastJet 3.3.4}~\cite{Cacciari:2011ma}, and the same reconstruction, isolation, and identification criteria are applied. Finally, we select events with at least three leptons, including a same-flavor oppositely charged lepton pair with invariant mass within the nominal $Z$-boson window and a third lepton consistent with originating from a $W$-boson (not from radiation). In addition, all further analysis-specific requirements, such as jet and $b$-tagged jet multiplicities and kinematic cuts on leptons and jets, are applied to ensure that the event samples match the signal regions defined by the CMS and ATLAS analyses. 

\section{Results and Interpretation}
The statistical model for the analysis is built from binned templates from data, SM predictions, and the NP contribution (see appendix for details). The NP signal strength is extracted via a simultaneous $\chi^2$ fit
\begin{align*}
\chi^2 = [\sigma_i^{\rm data} - \sigma_i^{\rm theory} ] \; \Sigma_{ij}^{-1} \; [\sigma_j^{\rm data} - \sigma_j^{\rm theory} ]\,,
\end{align*}
where $i,j$ run over the bins across all observables, $\Sigma_{ij}$ is the covariance matrix, $\sigma_i^{\rm data}$ is the measured cross section in bin $i$, and
\begin{align*}
\sigma_i^{\rm theory} = \mu_{\rm SM} \, \sigma_i^{\rm SM} + \mu_{\rm NP} \, \sigma_i^{\rm NP},
\end{align*}
represents the expected cross section in bin $i$, with SM and NP contributions weighted by fit parameters. Let us comment briefly on potential interference effects between the SM and NP contributions. While such effects can, in general, be relevant, the situation here differs qualitatively from the well-known heavy-Higgs interference in $gg\to t\bar t$~\cite{Bahl:2025you}, where resonant and continuum amplitudes depend on the same invariant-mass variable and the background develops a nontrivial complex phase across the resonance region. In the present case, the SM and NP contributions do not share a common resonant enhancement: the NP amplitude contains an intermediate on-shell $H^+$ in the decay chain, whereas the SM amplitude is smooth in the corresponding virtuality. Moreover, for $m_{H^+}<m_W+m_Z$ and with the OSSF dilepton selection enforcing an on-shell $Z$, the decay $H^+\to W^{(*)}Z$ necessarily proceeds through a strongly off-shell $W^*$, while the SM top decay produces an almost on-shell $W$. The resulting mismatch in resonant substructures and $W$-virtuality, together with the scalar--vector Lorentz-structure difference of the $H^+bt$ and $Wbt$ couplings and the small $\mathrm{Br}(t\to bH^+)$ in the parameter region of interest, provides multiple independent suppression mechanisms. Finally, the width with respect to the mass of our charged Higgs boson is at the sub-permille level, i.e.~it is extremely narrow. Therefore, the interference effects are negligible for the present analysis.

The correlations among the differential observables are obtained from our SM simulation. The NP signal strength $\mu_{\rm NP}$ is identified with ${\rm Br}(t \to H^\pm b)\times{\rm Br}(H^\pm \to W^\pm Z)$. For the CMS analysis, the theoretical uncertainty is included in the total uncertainty by adding it in quadrature to the experimental one, so we fix $\mu_{\rm SM} = 1$. In the ATLAS case, where theory uncertainties are not included in the reported errors, we use the one based on {\tt MG5\_aMC@NLO+Pythia 8}, whose predictions lie in between the two simulations obtained from {\tt SHERPA} (without and with mult-leg merging of additional partons). To account for this uncertainty, we profile over $\mu_{\rm SM}$, allowing a 5\% variation around 1, which corresponds to the uncertainty on the total $t\bar t$ production cross section~\cite{Czakon:2011xx}. 

\begin{figure*}[htb!]
\centering
\includegraphics[width=0.49\textwidth]{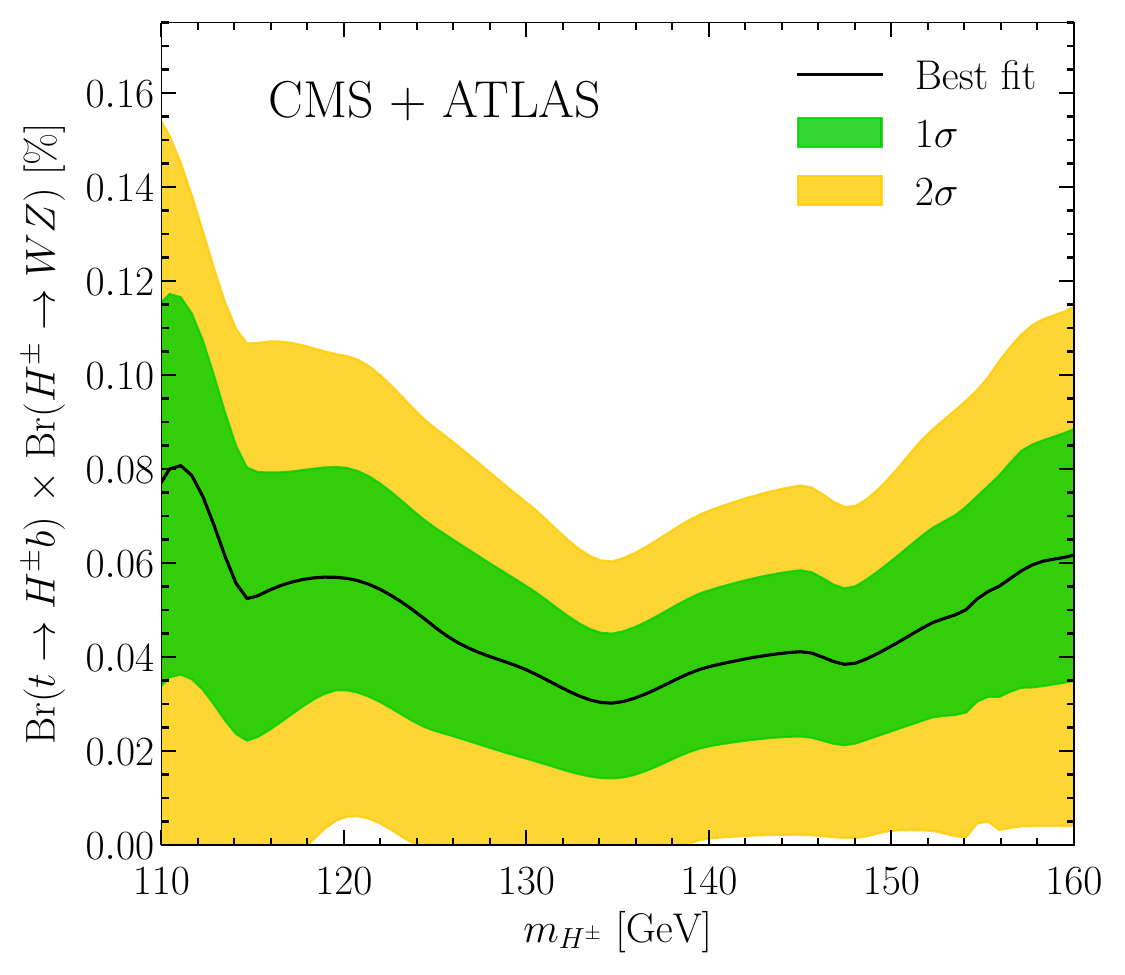}
\includegraphics[width=0.49\textwidth]{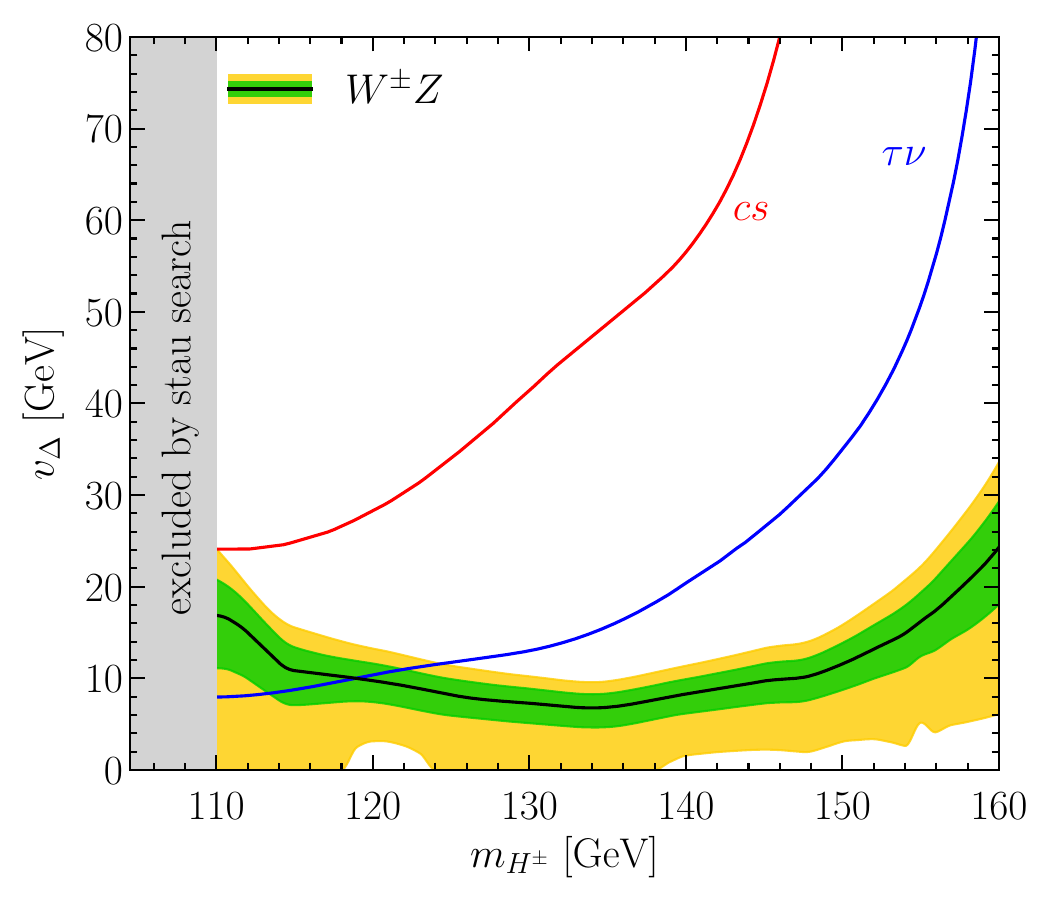}
\caption{Left: Preferred $1\sigma$ (green) and $2\sigma$ (yellow) range for ${\rm Br}(t \to H^\pm b) \times {\rm Br} (H^\pm \to W^\pm Z)$ as a function of $m_{H^\pm}$. Right: Preferred range for the $Y=0$ triplet VEV $v_\Delta$ from the $t\bar tZ$ measurements, interpreted within the gGM model, as a function of $m_{H^\pm}$. The gray band is excluded by stau searches, while the regions above the blue (red) line by LHC searches for $t\to H^\pm$ with $H^\pm b\to \tau\nu (cs)$ at 95\% CL.}
\label{fig:limit_combined}
\end{figure*}

Minimizing the global $\chi^2$ (ATLAS plus CMS), we extract model-independent bounds on ${\rm Br}(t\to H^\pm b) \times {\rm Br}(H^\pm\to WZ)$, with $\chi^2-\chi^2_{\rm min} \le 1~(4)$ defining the 1$\sigma$ (2$\sigma$) interval. The combined result is shown in the left panel of Fig.~\ref{fig:limit_combined} as a function of $m_{H^\pm}$ (the individual results for CMS and ATLAS are shown in Fig.~\ref{fig:limits} in the appendix). Note that the lower edge of the $2\sigma$ band lies very close to ${\rm Br}(t\to H^\pm b) \times {\rm Br}(H^\pm\to WZ)=0$, indicating a mild ($\sim 2\sigma$) preference for a NP contribution.

Next, we interpret these results within the GM model~\cite{Georgi:1985nv}, which contains a real $SU(2)_L$ triplet $\Delta$ with $Y=0$ and a complex triplet $\chi$ with $Y=1$. In its generic version (gGM)~\cite{Ashanujjaman:2025puu}, where custodial symmetry in the scalar potential is relaxed, the states originating predominantly from the $Y=0$ and $Y=1$ triplets can be non-degenerate in mass, even in the limit of small mixing between the two triplet sectors. This allows a light charged Higgs to be predominantly the charged component of the $Y=0$ triplet, $H^\pm\simeq H_\Delta^\pm\simeq\Delta^\pm$. In this limit, its phenomenology closely resembles that of the standalone real-triplet model, while the presence of the $Y=1$ triplet allows approximately equal triplet VEVs, $v_\chi\simeq v_\Delta$, consistent with electroweak precision data. Moreover, this sector remains largely unconstrained by LHC searches~\cite{Chabab:2018ert, Butterworth:2023rnw, Ashanujjaman:2024lnr}, so that its charged component can be lighter than the top quark, enabling the decay $t \to H^\pm b$ with the width
\begin{align*}
\Gamma(t \to H^\pm b) =& \frac{g^2_2}{64\pi m_W^2 m_t} \sin^2\beta |V_{tb}|^2 \lambda^{1/2} \left(\frac{m_b^2}{m_t^2}, \frac{m_{H^\pm}^2}{m_t^2}\right) \times
\\
& \left[ (m_t^2 + m_b^2 - m_{H^\pm}^2) (\bar m_t^2 + \bar m_b^2) + 4m_t^2 m_b^2 \right],
\end{align*}
where $\beta = \tan^{-1}(-2v_\Delta/v_\Phi)$ denotes the relevant charged Higgs mixing angle, with $v_\Phi$ being the SM Higgs VEV; $m_t = 172.5$ GeV and $m_b = 5.37$ GeV are the pole masses~\cite{ParticleDataGroup:2024cfk, Ma:2024xeq}, and $\bar m_b(m_t) \approx 2.6$ GeV is the $\overline{\rm MS}$ mass at scale $m_t$~\cite{Aparisi:2021tym} and $\lambda(x,y) = (1-x-y)^2-4xy$ is the usual kinematic function. While we provide the leading-order expression for $\Gamma(t \to H^\pm b)$, we include the NLO QCD correction following the $\mathcal{O}(\alpha_s)$ calculation of Ref.~\cite{Czarnecki:1992zm}.\footnote{The $H^\pm tb$ coupling in the $Y=0$ triplet model has the same chiral structure (including the finite $\epsilon = m_b/m_t$ terms) as the Model-I coupling of Ref.~\cite{Czarnecki:1992zm}. Since the QCD correction factor depends only on the chiral structure and kinematics, but not on the overall normalization, the result of Ref.~\cite{Czarnecki:1992zm} applies directly.} In our numerical implementation, we extract the correction factor from Fig.~1 of Ref.~\cite{Czarnecki:1992zm}, which provides $\delta_{\rm QCD} \equiv (\Gamma_{\rm NLO}-\Gamma_{\rm LO})/\Gamma_{\rm LO}$ as a function of $m_{H^\pm}/m_t$. For the mass range considered, the NLO QCD correction varies from approximately $-7\%$ at lower masses to about $-3\%$ around $m_{H^\pm}\sim150$~GeV and becomes positive close to threshold. This correction applies exclusively to $\Gamma(t \to H^\pm b)$. The SM top-quark width is taken at state-of-the-art accuracy, including NNLO QCD corrections~\cite{Czarnecki:1998qc, Chetyrkin:1999ju}, NLO electroweak corrections, and finite $b$-quark mass effects~\cite{Chen:2022wit}. For the reference value $m_t = 172.5$\,GeV, $\Gamma (t)_{\rm SM} = 1.326$\,GeV. In the parameter region relevant for our analysis, $\mathrm{Br}(t\to H^\pm b)\lesssim 10^{-3}$, so that the corresponding shift in the total top width remains at the per-mille level relative to $\Gamma(t)_{\rm SM}$.

In the right panel of Fig.~\ref{fig:limit_combined}, we show the limit on the $Y=0$ triplet VEV $v_\Delta$ obtained from our recast and compare it to those from searches for $t\to H^\pm b$ in the $cs$~\cite{ATLAS:2024oqu} (red) and $\tau\nu$~\cite{ATLAS:2024hya} (blue) modes. The recast of the stau search~\cite{CMS:2022syk} (gray) excludes charged Higgs masses below 110\,GeV~\cite{Ashanujjaman:2024lnr}. We see that our limits from the $WZ$ channel are stronger than those from the dedicated $cs$ and $\tau\nu$ searches. While such values of the triplet VEV would be strongly constrained in the standalone $Y=0$ triplet model, they remain compatible with electroweak precision data in the gGM model for approximately equal triplet VEVs, $v_\chi\simeq v_\Delta$.

\section{Conclusions and Outlook}
We have recast the latest LHC measurements of $t\bar t Z$ differential cross-sections to probe charged Higgs bosons produced in top-quark decays in the previously unexplored $H^\pm \to WZ$ decay mode. This results in stringent bounds on ${\rm Br}(t\to H^\pm b)\times{\rm Br}(H^\pm\to WZ)$ at the sub-permille level. Interpreted within the gGM model, these results probe the $Y=0$ triplet VEV at the ${\cal O}(10)$\,GeV level, providing stronger sensitivity than existing ATLAS searches in the $cs$ and $\tau\nu$ modes while remaining compatible with electroweak precision data.

Moreover, CMS and ATLAS data exhibit a combined $\sim\!\!2\sigma$ preference for a NP signal. While not significant in isolation, it further strengthens the cumulative case for a $152\pm 1$\,GeV boson, seen in di-photon and other measurements~\cite{Crivellin:2021ubm,Bhattacharya:2023lmu, Bhattacharya:2025rfr} (and predicated by the multi-lepton anomalies~\cite{vonBuddenbrock:2016rmr,vonBuddenbrock:2017gvy,Buddenbrock:2019tua,vonBuddenbrock:2020ter}), being the neutral component of an $SU(2)_L$ triplet with $Y=0$~\cite{Ashanujjaman:2024pky, Crivellin:2024uhc, Ashanujjaman:2024lnr}. In fact, in the gGM model the predominantly $Y=0$ charged and neutral states can remain close in mass, while the $Y=0$ triplet can play a crucial role in explaining the tensions in differential $t\bar t$ distributions~\cite{Banik:2023vxa,Coloretti:2023yyq}. Future high-luminosity LHC runs and dedicated analyses targeting the $H^\pm \to WZ$ mode in top decays could decisively test this scenario, potentially uncovering Higgs bosons beyond the SM.

\acknowledgments SA is supported by the Deutsche Forschungsgemeinschaft (DFG, German Research Foundation) under grant 396021762 - TRR 257. AC is supported by a professorship grant of the Swiss National Science Foundation (Grant No.\ PP00P21\_76884). SPM and BM acknowledge the support of the Research Office of the University of the Witwatersrand. BM further acknowledges support from the South African Department of Science and Innovation through the SA-CERN program, and the National Research Foundation. SA thanks Felix Yu for useful discussions.

\appendix
\section{}

The differential cross sections for the considered observables are shown in Fig.~\ref{fig:CMS_dist} and Fig.~\ref{fig:ATLAS_dist}, corresponding to the CMS and ATLAS analyses, respectively. The data and SM predictions are taken from each analysis, while the NP predictions correspond to $m_{H^\pm} = 150$\,GeV and to the respective best-fit values of ${\rm Br}(t \to H^\pm b) \times {\rm Br} (H^\pm \to W^\pm Z)$: 0.1\% for CMS and 0.04\% for ATLAS. Interestingly, the CMS measurement exhibits a small deviation from the SM prediction, while the ATLAS results are more consistent with the SM expectations.

The individual fits to ATLAS and CMS data are shown in Fig.~\ref{fig:limits}.

\clearpage

\begin{figure*}[htb!]
\centering
\includegraphics[width=0.95\textwidth]{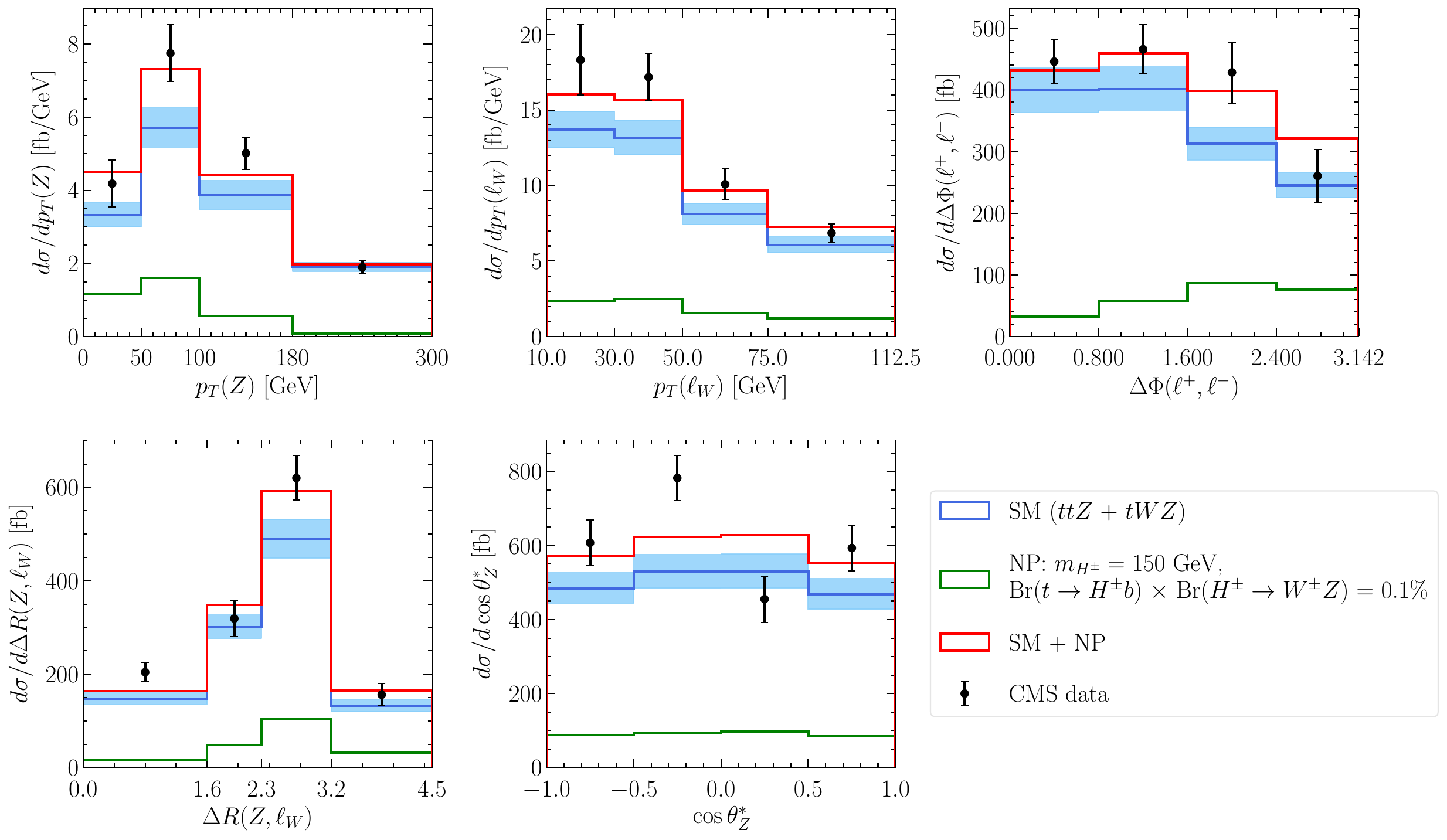}
\caption{The $ttZ+tWZ$ differential cross sections for different observables measured by CMS. The error bars indicate the total experimental uncertainties, while the shaded blue area corresponds to the uncertainty of the theory prediction. The NP contributions are shown for $m_{H^\pm} = 150$\,GeV and ${\rm Br}(t \to H^\pm b) \times {\rm Br} (H^\pm \to W^\pm Z) = 0.1\%$, corresponding to the best-fit to CMS data (see the left plot in Fig.~\ref{fig:limits}).}
\label{fig:CMS_dist}
\end{figure*}

\begin{figure*}[htb!]
\centering
\includegraphics[width=0.95\textwidth]{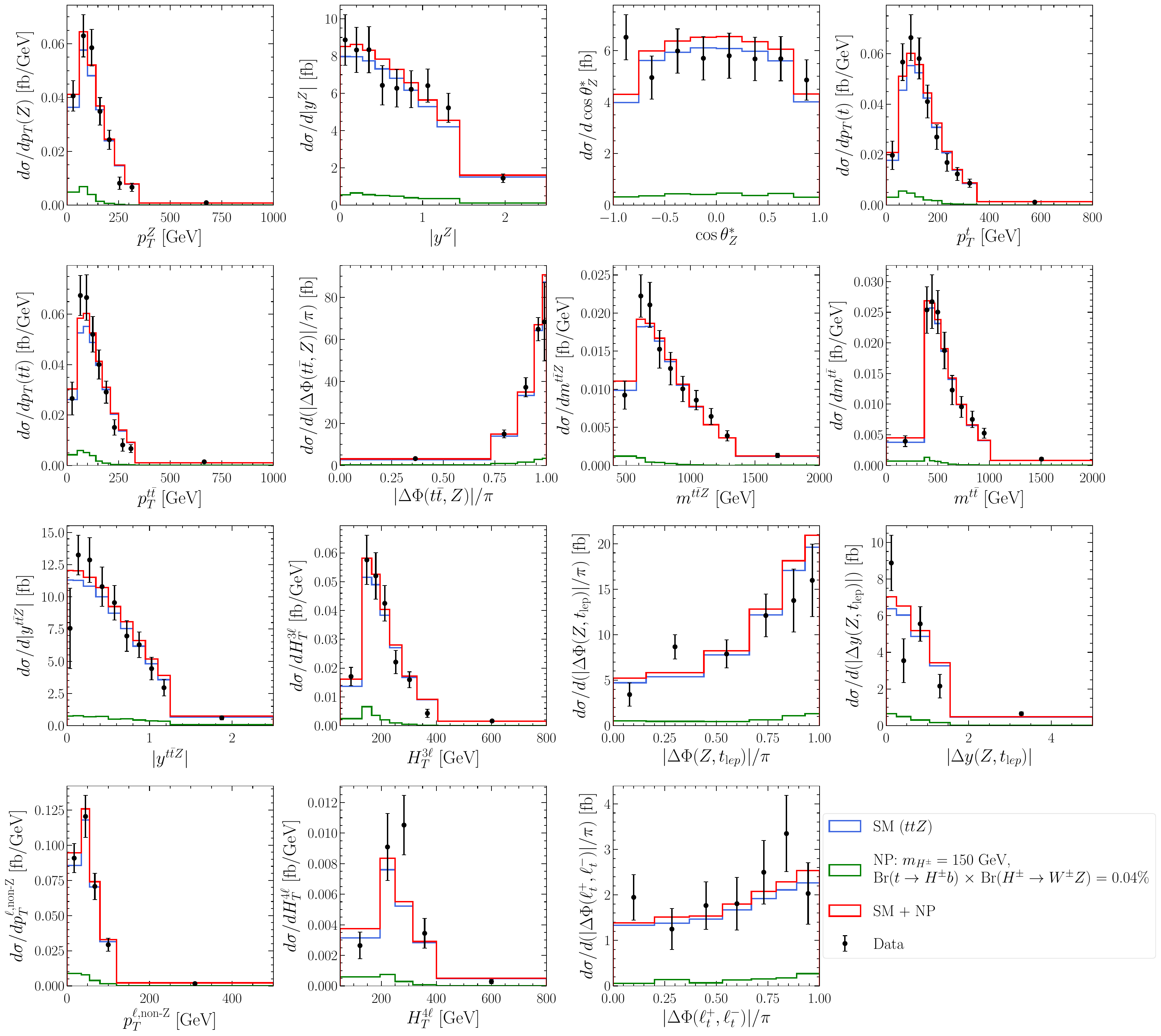}
\caption{Differential $ttZ$ cross sections unfolded to particle level for different observables measured by ATLAS. The error bars indicate the total experimental uncertainties. The NP contributions are shown for $m_{H^\pm} = 150$\,GeV and ${\rm Br}(t \to H^\pm b) \times {\rm Br} (H^\pm \to W^\pm Z) = 0.04\%$, corresponding to the best-fit to ATLAS data (see the right plot in Fig.~\ref{fig:limits}).}
\label{fig:ATLAS_dist}
\end{figure*}

\begin{figure*}[htb!]
\centering
\includegraphics[width=0.46\textwidth]{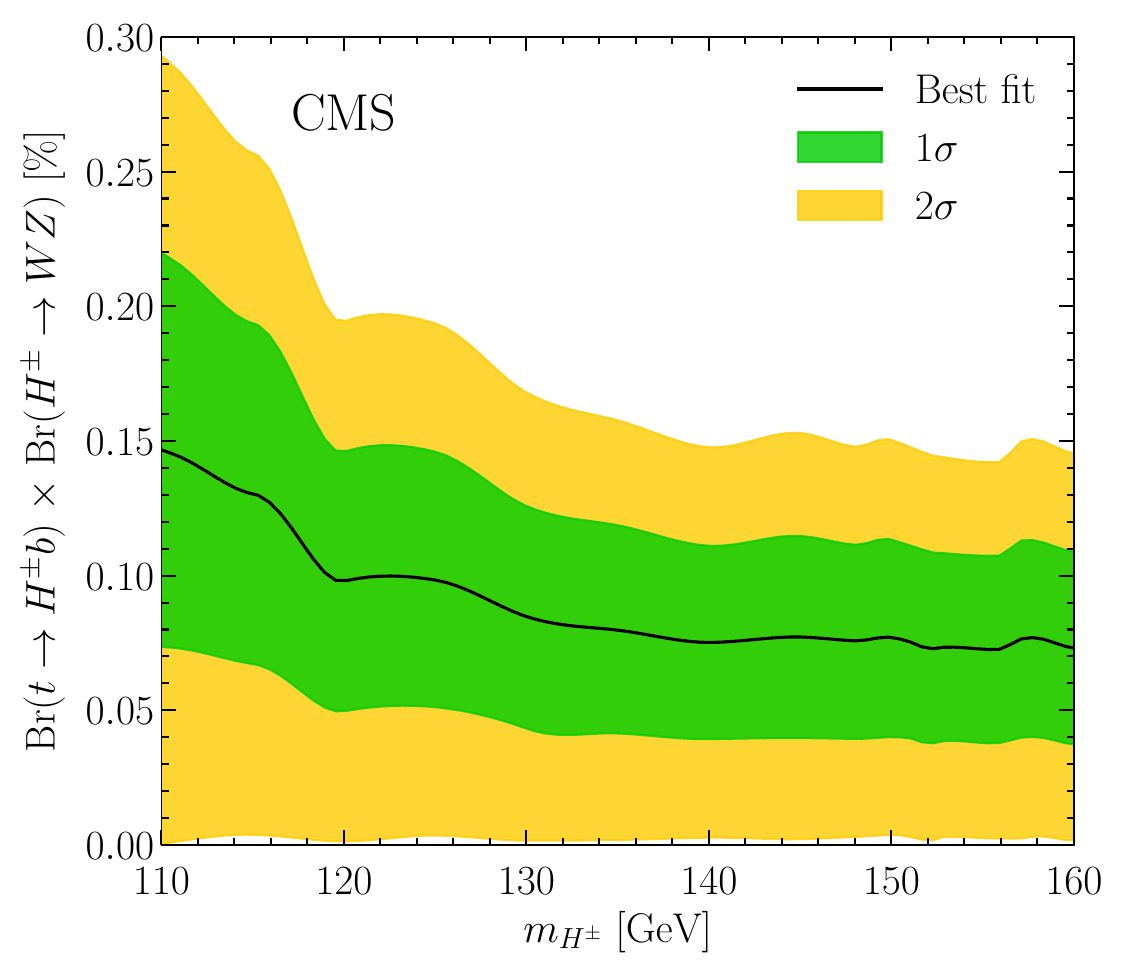}
\includegraphics[width=0.46\textwidth]{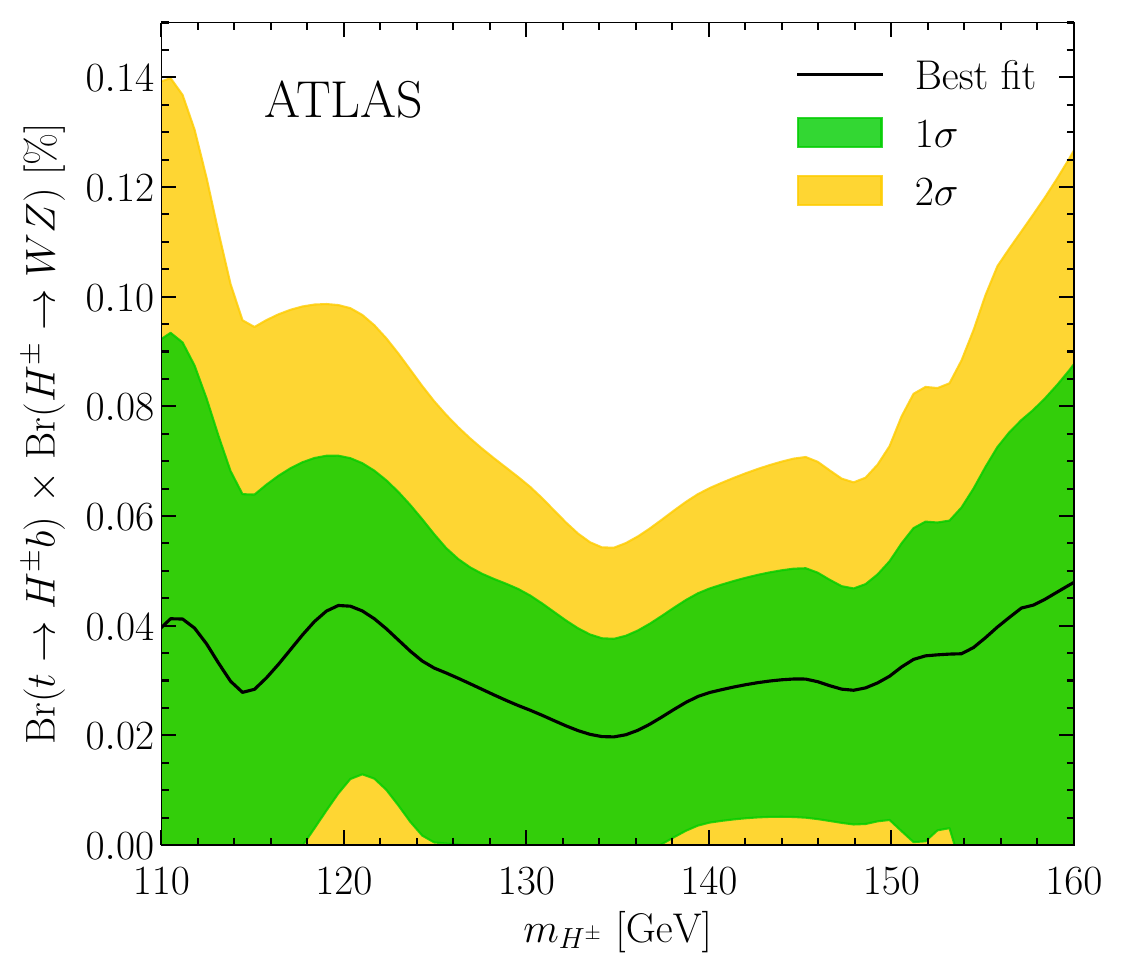}
\caption{Preferred $1\sigma$ (green) and $2\sigma$ (yellow) range for ${\rm Br}(t \to H^\pm b) \times {\rm Br} (H^\pm \to W^\pm Z)$ as a function of $m_{H^\pm}$, obtained from the CMS (left) and ATLAS (right) analyses.}
\label{fig:limits}
\end{figure*}

\bibliography{v0}

\end{document}